\documentclass[preprint2,numberedappendix]{emulateapj-rtx4}
\usepackage{graphicx,natbib}
\usepackage{bm,url,color}
\usepackage{amsmath}
\usepackage{amssymb}
\usepackage{graphicx}
\usepackage{euscript}
\graphicspath{{./figures/}{./png/}}
\graphicspath{{./fig/}{./png/}}

\newcommand{\EQ}{\begin{equation}}
\newcommand{\EN}{\end{equation}}
\newcommand{\EQA}{\begin{eqnarray}}
\newcommand{\ENA}{\end{eqnarray}}

\newcommand{\Eq}[1]{Eq.~(\ref{#1})}

\newcommand{\Sec}[1]{Sect.~\ref{#1}}

\newcommand{\Fig}[1]{Fig.~\ref{#1}}

\newcommand{\Tab}[1]{Table~\ref{#1}}

{}
{}
{}

{}
{}
{}
{}
{}
{}
{}
{}
{}
{}
{}
{}
{}
{}
{}
{}
{}
{}
{}
{}
{}
{}

{}
{}
{}

{}

{}
{}

\newcommand{\bb}{\bm{b}}
\newcommand{\BB}{\bm{B}}

\newcommand{\UU}{\bm{U}}
\newcommand{\uu}{\bm{u}}

\newcommand{\JJ}{\mbox{\boldmath $J$} {}}

\newcommand{\ff}{\mbox{\boldmath $f$} {}}

\newcommand{\EMF}{\mbox{\boldmath ${\cal E}$} {}}

\newcommand{\ii}{{\rm i}}

\newcommand{\const}{{\rm const}  {}}

\def\cp{c_{\rm p}}
\def\cv{c_{\rm v}}

\def\cs{c_{\rm s}}

\def\vA{v_{\rm A}}

\def\half{{\textstyle{1\over2}}}

\def\onethird{{\textstyle{1\over3}}}

\newcommand{\Mm}{\,{\rm Mm}}

\newcommand{\yapj}[3]{ #1, {ApJ,} {#2}, #3}

\newcommand{\yana}[3]{ #1, {A\&A,} {#2}, #3}

\newcommand{\ygafd}[3]{ #1, {Geophys.\ Astrophys.\ Fluid Dyn.,} {#2}, #3}

\newcommand{\yaraa}[3]{ #1, {ARA\&A,} {#2}, #3}
\newcommand{\yanf}[3]{ #1, {Ann. Rev. Fluid Mech.,} {#2}, #3}

\newcommand{\ysci}[3]{ #1, {Science,} {#2}, #3}
\newcommand{\ysph}[3]{ #1, {Solar Phys.,} {#2}, #3}

\newcommand{\ybook}[3]{ #1, {#2} (#3)}

\newcommand{\smn}[1]{ #1, {MNRAS}, submitted}
\newcommand{\beq}{\begin{equation}}
\newcommand{\eeq}{\end{equation}}

\newcommand{\Ratio}{q}

\newcommand{\re}{{\rm Re}}

\newcommand{\md}{{\rm Ma}}
\newcommand{\ul}{u_{\rm rms,d}}

\newcommand{\Lu}{L_{z\rm u}}
\newcommand{\Ld}{L_{z\rm d}}
\newcommand{\ru}{{\rho_{\rm u}}}
\newcommand{\rd}{{\rho_{\rm d}}}

\newcommand{\vaxd}{v_{{\rm A}x\rm d}}

\newcommand{\csu}{c_{\rm su}}
\newcommand{\csd}{c_{\rm sd}}

\newcommand{\kxB}{k^B_x}

\shorttitle{\uppercase{Fanning out of the $f$-mode}}
\shortauthors{\uppercase{Singh, Brandenburg, \& Rheinhardt}}

\begin{document}

\title{Fanning out of the $f$-mode in presence of nonuniform magnetic fields}
\author{Nishant K. Singh\altaffilmark{1},
Axel Brandenburg\altaffilmark{1,2}, and
Matthias Rheinhardt\altaffilmark{3}}
\email{nishant@nordita.org}
\altaffiltext{1}{Nordita, KTH Royal Institute of Technology and Stockholm University,
Roslagstullsbacken 23, SE-10691 Stockholm, Sweden}
\altaffiltext{2}{Department of Astronomy, Stockholm University, SE-10691 Stockholm, Sweden}
\altaffiltext{3}{Physics Department, Gustaf H\"allstr\"omin katu 2a, PO Box 64,
FI-00014 University of Helsinki, Finland}

\date{\today,~ $ $Revision: 1.119 $ $}

\begin{abstract}
We show that in the presence of a harmonically varying magnetic field
the fundamental or $f$-mode in a stratified layer is altered in such
a way that it fans out in the diagnostic
$k\omega$ diagram, but with mode power also within the fan.
In our simulations, the surface is defined by a temperature and density jump
in a piecewise isothermal layer.
Unlike our previous work \citep{SBCR14} where a uniform magnetic field
was considered, we employ here a nonuniform magnetic field together with
hydromagnetic turbulence at length scales much smaller than those
of the magnetic fields.
The expansion of the $f$-mode is stronger for fields confined to the
layer below the surface.
In some of those cases, the $k\omega$ diagram also reveals a new class
of low frequency vertical stripes at multiples of twice the horizontal
wavenumber of the background magnetic field.
We argue that the study of the $f$-mode expansion might be a new and
sensitive tool to determining subsurface magnetic fields with
longitudinal periodicity.
\end{abstract}
\maketitle

\section{Introduction}
\label{Intro}

For several decades, helioseismology has provided information about the
solar interior through detailed investigations of sound or pressure waves,
generally referred to as $p$-modes.
While internal gravity waves or $g$-modes,
are evanescent in the convection zone and hence not seen in the Sun,
the so-called surface or fundamental mode ($f$-mode) is observable.
This mode is just like deep water waves.
In that case it is well known that the presence of surface tension leads to
additional modes known as capillary waves \citep{DK99}.
Such modes do not exist on gaseous interfaces, but magnetic fields could
mimic the effects of surface tension and thus lead to characteristic
alterations of the $f$-mode, that could potentially be used to determine
properties of the underlying magnetic field.
Earlier work has indeed
shown that both vertical and horizontal uniform magnetic fields
have a strong effect on the $f$-mode \citep[][hereafter SBCR]{SBCR14}.
But obviously, the assumption of a uniform magnetic field is unrealistic.

The goal of local helioseismology using the $f$-mode
\citep{HBBG08,DABCG11,FBCB12,FCB13} is to determine the local structure
of the underlying magnetic field.
Such techniques might be more
sensitive than local techniques employing just $p$-modes
\citep[see, e.g.,][]{GBS10}.
The goal here is to determine the structure of sunspot magnetic fields
and to decide whether they have emerged as isolated
flux tubes from deeper layers \citep{Par75,Cal95},
as expected from the flux transport dynamo paradigm.
An alternative approach to solar magnetism presumes that
the dynamo is a distributed one
operating throughout the entire convection zone and not just at 
its bottom,
and that sunspots are merely localized flux concentrations near the surface.
This approach was discussed in some detail by \cite{B05}, who mentioned the
negative effective magnetic pressure instability \citep{KMR96}
and the local suppression of turbulent heat transport \cite{KM00}
as possible agents facilitating the formation of 
such magnetic flux concentrations.
He also discussed magnetic flux segregation into magnetized and unmagnetized
regions \citep{TWBP98} as a mechanism involved in the formation of active
regions.
However, in those two instabilities, the magnetic field experiences an
instability that leads to field concentrations locally near the surface.
In the horizontal plane, the field shows a periodic pattern that also
plays a role in the motivation of the field pattern chosen for the
present investigation.

\section{Model setup and motivation}

Our model is similar to that studied in SBCR, where
we adopt a piecewise isothermal setup with a lower cooler layer
(`bulk' with thickness $L_{z\rm d}$) and a hotter upper one
(`corona' with thickness $L_{z\rm u}$).
We solve the basic hydromagnetic equations,
\begin{align}
\frac{D \ln \rho}{Dt} &= -\bm\nabla\cdot\bm{u}, \\
\frac{D\bm{u}}{Dt} &= \ff+\bm{g} +\frac{1}{\rho}
\left(\bm{J}\times\bm{B}-\bm\nabla p
+\bm\nabla \cdot 2\nu\rho\bm{\mathsf{S}}\right),\\
T\frac{D s}{Dt} &= 2\nu \bm{\mathsf{S}}^2 + \frac{\mu_0\eta}{\rho}\JJ^2
- \gamma (c_p-c_v) \frac{T-T_{\rm d,u}}{\tau_{\rm c}}\, ,
\label{equ:ss} \\
\frac{\partial \bm A}{\partial t} &= {\bm u}\times{\bm B} 
+ \EMF_0- \eta \mu_0 {\bm J},
\end{align}
where $\bm{u}$ is the velocity,
$D/Dt = \partial/\partial t + \bm{u} \cdot \bm\nabla$ is the
advective time derivative,
$\bm{f}$ is a forcing function specified below,
$\bm{g}=(0,0,-g)$ is the gravitational acceleration,
$\mathsf{S}_{ij}=\half(u_{i,j}+u_{j,i})
-\onethird \delta_{ij}\bm\nabla\cdot\bm{u}$
is the traceless rate of strain tensor, where commas denote
partial differentiation,
$\nu=\const$ is the kinematic viscosity,
${\bm A}$ is the magnetic vector potential,
${\bm B} =\bm\nabla\times{\bm A}$ is the magnetic field,
${\bm J} =\mu_0^{-1}\bm\nabla\times{\bm B}$ is the current density,
$\EMF_0$ is an external electromotive force specified below,
$\eta=\const$ is the magnetic diffusivity, $\mu_0$ is the vacuum
permeability, and $T$ is the temperature.
The last term in \Eq{equ:ss}, being of relaxation type, is
to guarantee that the temperature is on average constant in
either subdomain and equal to $T_{\rm d}$ and $T_{\rm u}$, respectively.
For the relaxation rate $\tau_{\rm c}^{-1}$ we choose
$0.5\,g/\csd$ in $z>0$ and, for simplicity, zero in $z<0$
throughout this paper.

The adiabatic sound speeds in the upper and lower layers
are referred to as $\csu$ and $\csd$, respectively.
In most of the cases, we assume a temperature jump 
$\Ratio =T_{\rm u}/T_{\rm d}= \csu^2/\csd^2=\ru(0)/\rd(0)$
of about one tenth, which means that at the interface the density
changes by the same factor, allowing thus for the $f$-mode to appear.
A random flow is driven in the lower layer ($z<0$) by applying a solenoidal
non-helical forcing with a wavenumber that is much larger than the lowest
one fitting into the domain.
We normalize the length scales by $L_0= \gamma H_{\rm d}=\csd^2/g$,
where $\gamma=\cp/\cv$ is the ratio of specific heats
at constant pressure and density, respectively, and
$H_{\rm d}$ is the pressure scale height in the bulk.
Frequencies are normalized by $\omega_0=g/\csd$
and quantities normalized this way are indicated by tildae.

In a customary $k_x$-$\omega$ diagram (referred to simply as $k\omega$ diagram),
we show the amplitude of the Fourier transform of the vertical velocity
$u_z$, taken from the interface at $z=0$, as a function of $k_x$ and
$\omega$. As its Fourier transform $\hat{u}_z(k_x, \omega)$ has the
dimension of length squared, we construct the dimensionless quantity
\EQ
\widetilde{P}(\omega, k_x)\;=\;\frac{|\hat{u}_z|}{{\cal D}^2}
\;=\; \frac{|\hat{u}_z|}{L_0^2}\,\frac{\csd^2}{\ul^2},
\label{P}
\EN
where $\ul$ represents the root-mean-squared value of turbulent
motions in the bulk and ${\cal D}=\ul/\omega_0$ 
is the distance traveled with speed $\ul$ in a time $\omega_0^{-1}$.
The fluid Reynolds and Mach numbers of the flow are
defined as $\re=\ul/(\nu k_{\rm f})$ and $\md=\ul/\csd$, respectively.

As in SBCR, we employ a two-dimensional model in $x$ and $z$,
ignoring any variations in the $y$ direction.
The domain is of size $L_x\times L_z = 8\pi L_0\times\pi L_0$,
where $L_x$ and $L_z$ denote the horizontal and vertical extent, respectively.
For the boundary conditions 
we adopt a perfect conductor and vanishing stress
at top and bottom of the domain
and periodicity in $x$ direction.

\begin{figure}
\begin{center}
\includegraphics[width=\columnwidth]{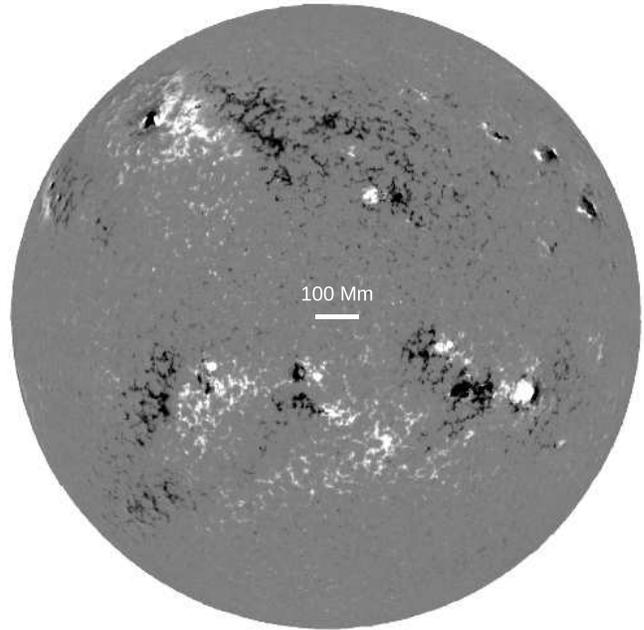}
\end{center}
\caption[]{
Full disk solar magnetogram showing the line of sight magnetic field
during a very active phase.
Light (dark) shades correspond to positive (negative) values.
Note the characteristic wavelength $\approx 100\Mm$ of regular
sign changes.
}\label{SM}
\end{figure}

\begin{figure*}
\begin{center}
\includegraphics[scale=1.]{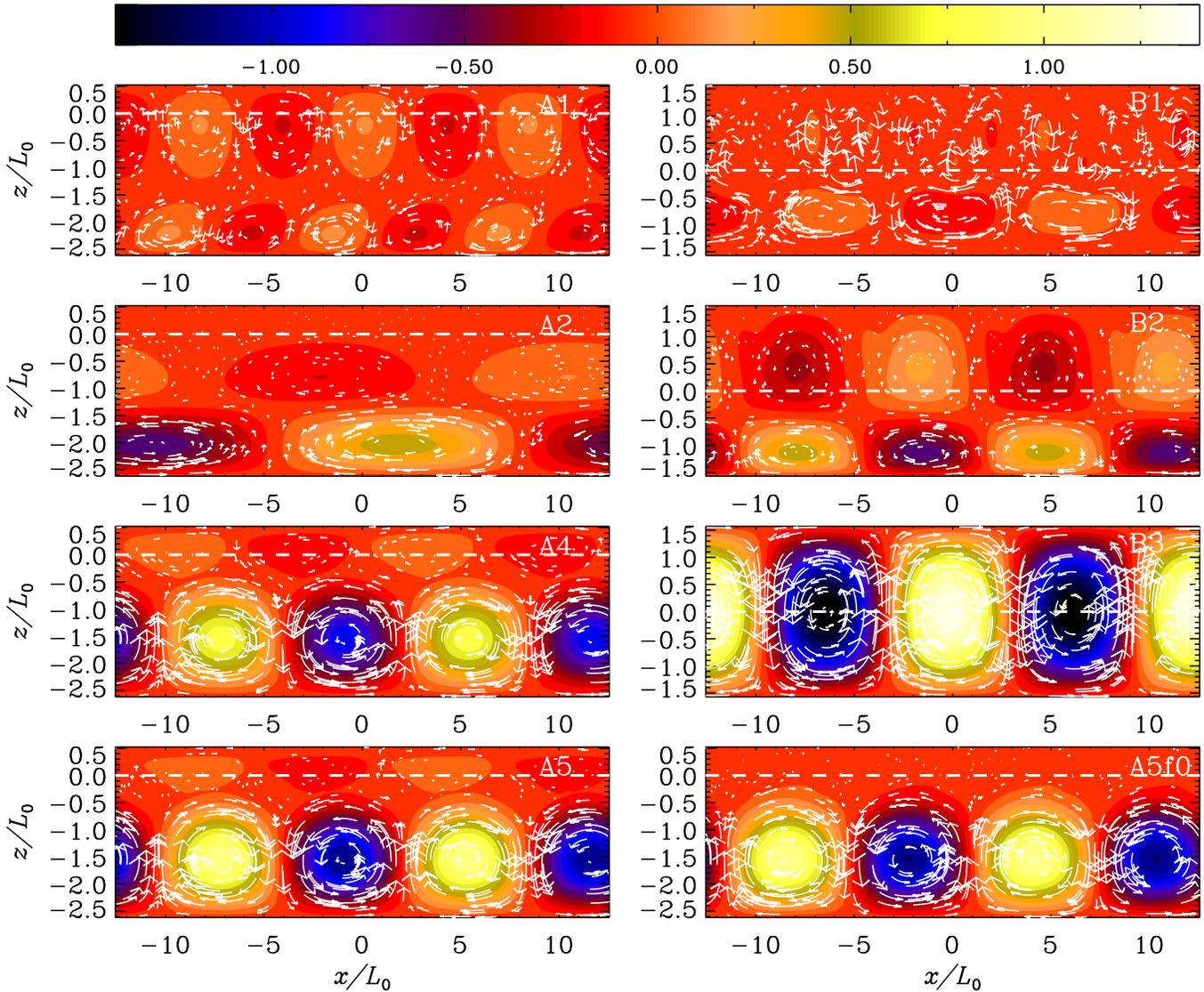}
\end{center}
\caption[]{
Saturated magnetic background fields of the runs in
\Tab{table_runs}.
Colors and arrows indicate the vector potential $A_y$ and the
magnetic field ${\bm B}$, respectively.
}\label{mag_str_last}
\end{figure*}

As discussed at the end of \Sec{Intro}, we produce
a steady magnetic field $\BB_0$ by applying a constant external        
electromotive force $\EMF_0$ with a harmonic spatial variation
\begin{equation}
{\cal E}_{0y}=\hat{\cal E}_0 \cos (k^B_x x) \cos (k^B_z z).
\end{equation}
Our choice of a sinusoidally varying magnetic field can be motivated
by looking at a solar magnetogram showing a regular pattern of alternating
positive and negative vertical field along the azimuthal direction (\Fig{SM}).
However, the stationary magnetic field that emerges
in the domain is the result of the combined effects of $\EMF_0$ and
the Lorentz force of $\BB_0$.
Given that initially, when the fluid is still at rest,
$\EMF_0$ generates a field resembling
that of an array of (thick) straight wires in the $y$ direction,
the Lorentz force will tend to compress these field vortices (rolls)
and hence accumulate fluid within them.
Consequently, they have to sink to a position where
their excess weight is compensated by the magnetic pressure gradient
of the field concentrated between the rolls and the lower boundary.
Finally, the overall state adjusts to a steady MHD equilibrium
$(\UU_0, \rho_0, \BB_0)$ which can qualitatively be characterized
by the spectra of its fields with respect to $x$.
Assuming $\kxB= N k_1$, where $k_1 = 2\pi/L_x$ (here, $k_1 \gamma H_{\rm d}=0.25$), the
spectra of $\UU_0$ and $\rho_0$ are given by
$\pm 2 m N k_1$ and that of $\BB_0$ by $\pm(2m+1) N k_1$ with $m=0,1,2,\ldots$.
In \Fig{mag_str_last} we show visualizations of the background field $\BB_0$ 
for all runs discussed in the results section; see \Tab{table_runs}.

We define the $z$ dependent root-mean-squared
Alfv\'en speed, $v_{\rm A}(z)$, and the quantity $\beta$,
characterizing the subsurface concentration of $v_{\rm A}(z)$, as
\EQ
v_{\rm A}(z) = \sqrt{\frac{\langle B^2 \rangle_x(z)}{\mu_0 \rho(z)}}\;,\quad
\beta=\big(\max_{z\leq0}\,{v_{\rm A}(z)}\big)/{c_{\rm sd}}.
\label{vaz}
\EN
In \Fig{va_cs_z} we show the variation of 
$v_{\rm A}(z)/c_{\rm s}(z)$ with $z$
for all runs; more details are given in \Tab{table_runs},
where $z_{\rm m}$ is the position of the maximum in \Eq{vaz}.
We also show that the $f$-mode asymmetry, characterized by the quantity
${\cal A}_{\rm f}$ (defined below) increases with $z_{\rm m}$
and varies only weakly with $\beta$.

For a horizontally imposed uniform magnetic field,
the $f$-mode frequency is well described
by the dispersion relation \citep{C61,MR92,MAR92}
\EQ
\omega_{\rm fm}^2 = c_{\rm fm}^2 k_x^2 +
g k_x \frac{1-\Ratio}{1+\Ratio},
\label{DR-SHM}
\EN
where $c_{\rm fm}^2=2\rd \vaxd^2/(\rd+\ru)$ with
$\vaxd$ being the Alfv\'en speed just below the interface;
see Equation~(21) of SBCR.
Here, the second term on the right-hand side represents the
(square of the) classical, unmagnetized $f$-mode frequency, to which
the first term, being the magnetic contribution, always adds.
Thus, horizontal magnetic fields lead to an increase in the $f$-mode frequency,
but, as discussed by \cite{MR93a},
turbulence without magnetic field leads to a decrease.
SBCR found that also strong {\em vertical} magnetic fields
lead to a decrease
of the $f$-mode frequency for sufficiently large $\widetilde{k}_x$. 

\cite{BH87} analyzed the alterations of the $p$-mode frequencies
in the presence of a nonuniform (piecewise uniform periodic)
magnetic field, but they did not consider $f$-modes.
It would be important to determine how a harmonic magnetic field
affects the $f$-mode frequencies, but such calculations
have not yet been done.
Some qualitative insight can be gained from an analysis of the possible eigensolution spectra.
In the linearized MHD equations, the coefficients 
of the perturbations $\uu$, $\rho'$, and $\bb$,
which are essentially determined by the background
fields $\UU_0$, $\rho_0$, and $\BB_0$,
are periodic in $x$ (or constant).
Hence, the eigenmodes must in general comprise an
infinitude of $x$ wavenumbers.
With the spectra of the background fields derived
above we expect for $\uu$ and $\rho'$ non-vanishing
spectral amplitudes at $(\pm2 m N  \pm l ) k_1$, but
for $\bb$ at  $\big(\pm(2m+1) N \pm l \big) k_1$,
where $m=0,1,2,\ldots$, and $l$ is a fixed
integer, $0\le l \le N/2$.

The described eigenmodes correspond to Bloch waves
being bounded solutions of the stationary Schr\"odinger
equation with a periodic potential.
According to Bloch's theorem they must have the
form $c_+ F(x) \exp \ii k_0 x   + c_- F(-x) \exp (-\ii k_0 x) $,
where $F(x)$ is a function with the same periodicity
as the potential and $k_0$ is the so-called Bloch wavenumber \citep{BH87}.
For our conditions, $k_0/k_1=l$ and can hence only adopt integers from
0 to $N/2$.
\\

\begin{table}\caption{
Summary of simulations with 
$\Ratio=0.1$, $\nu/\eta=1$ and $\widetilde{k}_{\rm f}=20$.
$\Delta{\omega}_{\rm f\pm}/{\omega}_{\rm f}$ are evaluated at
$\widetilde{k}_x=9$.
}\centering
\label{table_runs}
\resizebox{\columnwidth}{!}{%
\begin{tabular}{l l r c c c c c c }
\hline\hline\\[-2.5mm]
\hspace*{-1mm}Run & $\widetilde{k}^B_x$ & $\widetilde{k}^B_z$ & $\beta$ & $\widetilde{z}_{\rm m}$
& $\frac{\Delta{\omega}_{\rm f+}}{{\omega}_{\rm f}}$ & $\frac{\Delta{\omega}_{\rm f-}}{{\omega}_{\rm f}}$
& $\re$ & $\md$ \\ [.5mm]
\hline\\[-3mm]
A1 & 0.75 & 2.5 & 0.074 & 0 & - & - & 4.95 & 0.0198 \\
A2 & 0.25 & 3.0 & 0.118 & -1.5 & 0.33 & 0.23 & 9.67 & 0.0201 \\
A4 & 0.50 & 2.0 & 0.234 & -0.8 & 0.33 & 0.37 & 0.79 & 0.0032 \\
A5 & 0.50 & 2.0 & 0.285 & -0.8 & 0.44 & 0.41 & 1.15 & 0.0046 \\
A5f0\footnote[$\dagger$]{\scriptsize no random forcing} &0.50 & 2.0 & 0.278 & -0.75 & 0.11 & 0.0 & 0.74 & 0.0030 \\
B1 & 0.50 & 4.0 & 0.034 & -0.3 & - & - & 0.05 & 0.0010 \\
B2 & 0.50 & 2.0 & 0.104 & -0.7 & 0.33 & 0.30 & 1.06 & 0.0042 \\
B3 & 0.50 & 1.0 & 0.127 & 0 & - & - & 0.03 & 0.0006 \\
\hline
\end{tabular}%
}\end{table}

\section{Results}

To demonstrate the effects of nonuniformity of the magnetic field,
we study two types of cases: for the first one
the domain is asymmetric with respect to the 
interface ($L_{z,\rm d}/L_{z,\rm u}= 5$; Runs~A1--A5),
while symmetric for the second ($L_{z,\rm d}=L_{z,\rm u}$; Runs~B1--B3).
 
\begin{figure}
\begin{center}
\includegraphics[width=\columnwidth]{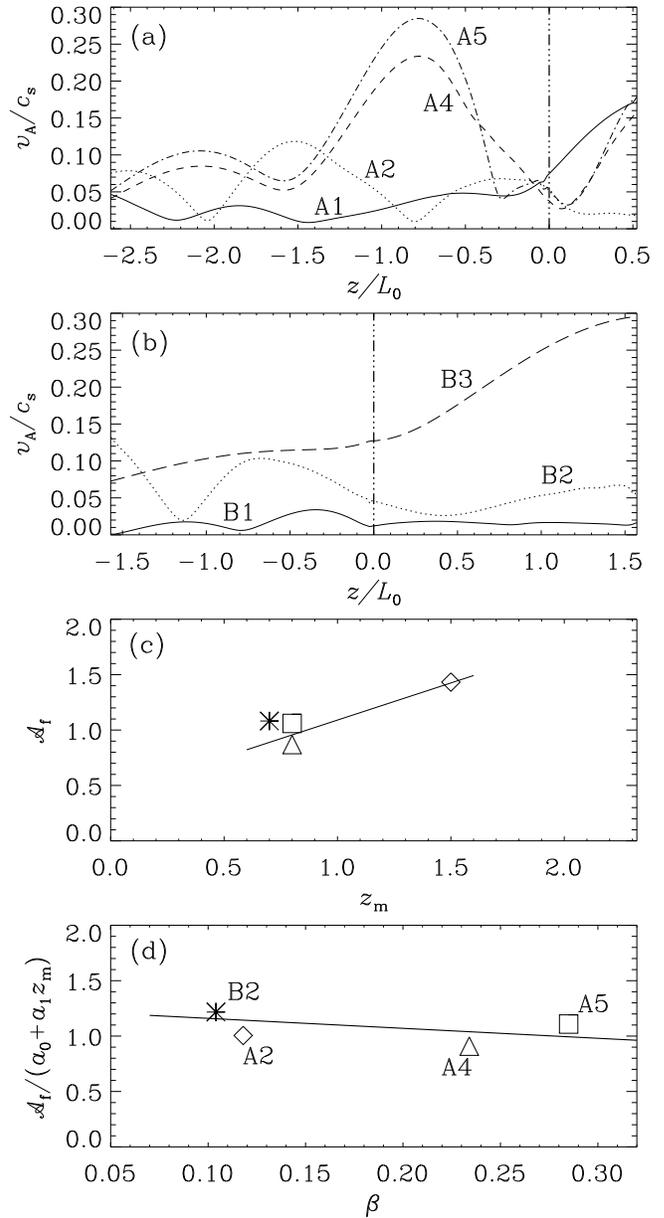}
\end{center}
\caption[]{
Vertical profiles of the ratio of Alfv\'en to sound speed, $\vA/\cs$
in upper two panels; dash-dotted line: position of the interface.
${\cal A}_{\rm f}$ as a function of $z_{\rm m}$ in panel (c)
showing the fit ${\cal A}_{\rm f}=a_0+a_1z_{\rm m}$
with $a_0=0.42$ and $a_1=0.67$ for all values of $\beta$ (solid line);
${\cal A}_{\rm f}$ normalized by this fit as
a function of $\beta$ in panel (d).
}\label{va_cs_z}
\end{figure}

\begin{figure*}
\begin{center}
\includegraphics[scale=1.]{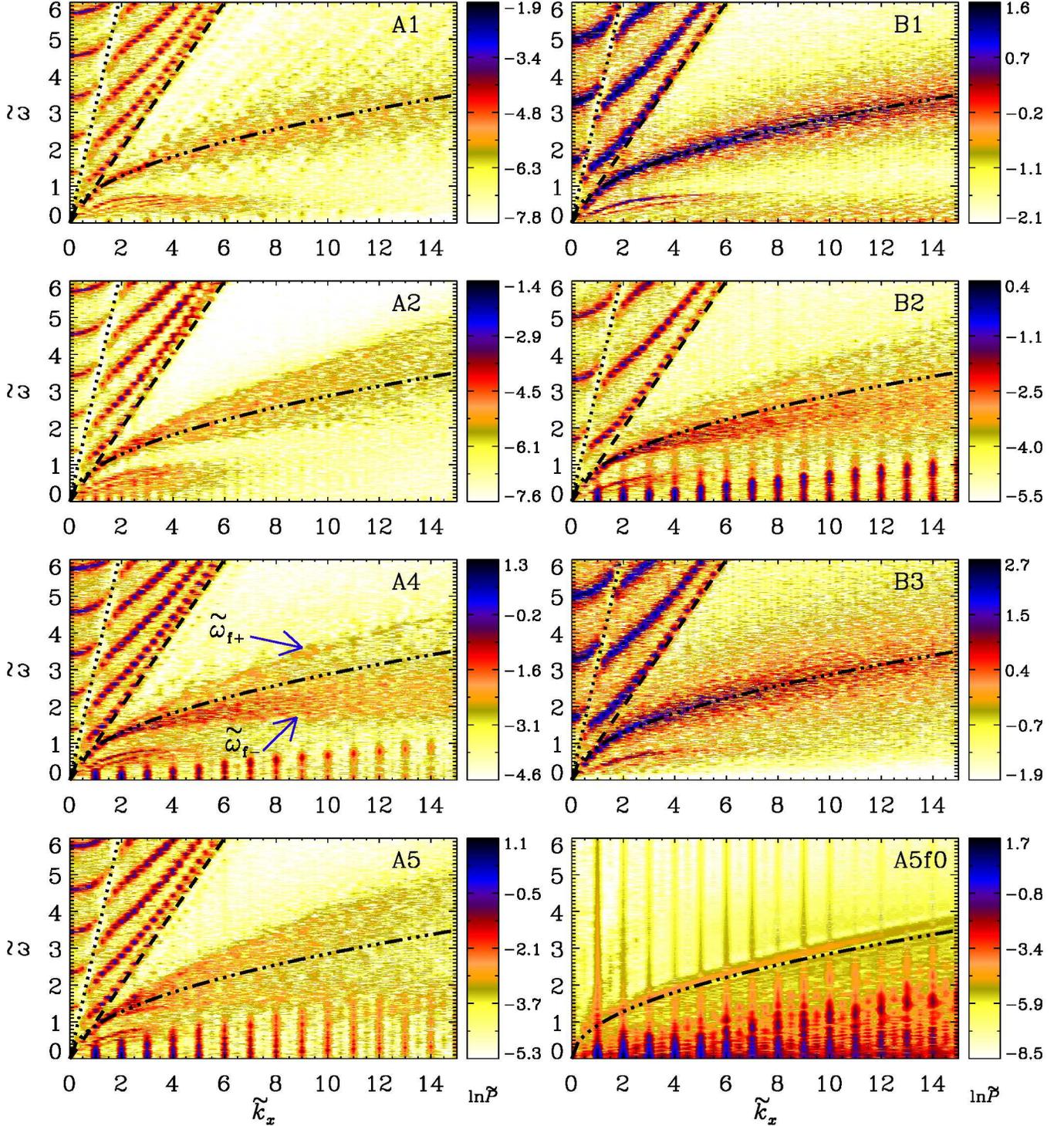}
\end{center}
\caption[]{$k\omega$ diagrams corresponding to the runs
of \Tab{table_runs}; $\Ld/\Lu=5$ for models
A1--5 and $\Ld=\Lu$ for models B1--3.
The dotted and dashed lines show $\omega=\csu k_x$ and $\omega=\csd k_x$,
respectively; dash-dotted curves show $\omega_{\rm f}$ (the classical $f$-mode).
In panel A4, the arrows indicate the estimated edges of the $f$-mode
fan at $\widetilde{k}_x=9$.
}\label{ko}
\end{figure*}

The corresponding $k\omega$ diagrams are shown in \Fig{ko}.
Similar to the nonmagnetic or weakly magnetized cases studied by SBCR,
we see $p$-modes above the line $\omega=\csd k_x$ with an apparent
discontinuity at $\omega=\csu k_x$, and indications of $g$-modes at
$\widetilde{k}_x<4$ and $\widetilde\omega=0.5$--$0.8$.
However, the $f$-mode now fans out and spans a trumpet-shaped
structure around the non-magnetic $f$-mode frequency $\sqrt{gk}$.
The more the magnetic field is pushed toward the bottom
of the domain, the more asymmetric this expansion appears to be with respect
to the usual $f$-mode in the unmagnetized case.

To quantify the fanning out of the $f$-mode, we denote 
the upper and lower edges of the fan at a given value of ${k}_x$ 
by ${\omega}_{\rm f+}$ and ${\omega}_{\rm f-}$, respectively.
In general, the fan is asymmetric
with respect to the classical $f$-mode ($\omega_{\rm f}$).
Let us represent this asymmetry at any given ${k}_x$ by
${\cal A}_{\rm f}=\Delta {\omega}_{\rm f+}/\Delta {\omega}_{\rm f-}$,
where $\Delta {\omega}_{\rm f\pm}=| \omega_{\rm f\pm}-\omega_{\rm f}|$
are the frequency spreads above and below
$\omega_{\rm f}$; see \Tab{table_runs} for
$\Delta {\omega}_{\rm f\pm}/\omega_{\rm f}$ at $\widetilde{k}_x=9$.
We find that at $\widetilde{k}_x=15$, the total relative spread,
$(\Delta {\omega}_{\rm f+}+\Delta {\omega}_{\rm f-})/\omega_{\rm f}$,
can be as large as $1.1$, which further increases with increasing
$\beta$ for fixed $z_{\rm m}$; see Runs~A4 and A5 in \Fig{ko}.
Note that ${\cal A}_{\rm f}$ can take values both
larger and smaller than unity, as may be seen by comparing Run~A2
with Runs~A4 or A5 in \Fig{ko}; see also \Tab{table_runs} and \Fig{va_cs_z}.

In addition, we see as a qualitatively new feature
a regular pattern of vertical stripes at multiples
of $2\widetilde{k}_x^B$ all the way up to $\widetilde{k}_x=14$, which appears
unconnected with the $f$-mode.
(In the spectra of $\BB$ the stripes appear at odd multiples of $\widetilde{k}_x^B$.)
They are absent if $\BB_0$ is independent of $z$
(${k}^B_z=0$; not shown), but most pronounced when
the magnetic field is concentrated in the lower part of the domain.
These are also cases in which the $f$-mode appears most fanned out.
The stripes appear weaker when the magnetic field is symmetric about the 
interface at $z=0$ (Run~B3) or when it is generally weak
(Runs~A1, A2, and B1).
Given that they are persistent after switching off the random
hydrodynamic forcing (see Run~A5f0 in \Fig{ko}), they can be identified 
to indicate at least one {\em unstable} eigensolution. Note that for a fixed $\omega$ an infinitude of  $k_x$ belongs to the 
same eigenmode. The discrete $\omega$ spots within each stripe may
either   belong to different unstable eigenmodes or represent overtones of a single mode.
The velocity field of the stripes is close to solenoidal and their 
occurrence and amplitude are strongly dependent on the strength of $\BB_0$. 
So we propose to consider
them as shear Alfv\'en modes having become unstable due to the inhomogeneity of $\BB_0$.
A similar transition is observed in whistler waves which become unstable
for suitably non-uniform background fields \citep{RG02}.

Remarkably, the unstable mode(s) excite $f$ modes, but without fanning
them out, whereas $p$ modes remain unexcited (Panel A5f0).
Comparing panels A5 and A5f0 suggests that
the fanning out requires not just a non-uniform
magnetic field, but also the presence of random forcing.
However, the fact that A5f0 exhibits only a single line and no fan might
indicate a physical difference between the fan and the regular $f$-mode.
We also note that the expansion of the $f$-mode
can still be seen when the vertical stripes
are weak or absent (especially in Run~A2), but in Run~B1, where
the field is less deep and the domain symmetric about $z=0$,
the $f$-mode lacks a clear trumpet shape.

\section{Conclusions}

The present study was aimed at identifying diagnostic signatures of
spatial variability of the magnetic field.
Indeed, we find in
the fanning out of the $f$-mode and in a pattern of vertical stripes in the $k\omega$ diagram such characteristic
features which have not been reported in earlier helioseismic studies.

The fanning out of the $f$-mode is different from the case of capillary
waves, where instead a ``bifurcation'' of the $f$-mode in deep water
waves is caused by surface tension \citep{DK99}.
In the present case, the width of the fan and its asymmetry
appear to characterize the magnetic field strength.
Independent from that, the horizontal variability
of the underlying magnetic field
is reflected in the presence of vertical stripes in the diagnostic
$k\omega$ diagram at even multiples of the horizontal wavenumber of the
magnetic field.
We have proven that the stripes can be assigned to one or perhaps several
unstable eigenmodes, most likely of shear Alfv\'en type.

The spatial variation of the photospheric field
seen in \Fig{SM} with a wavelength of about $100\Mm$
corresponds to a wavenumber $k_x=0.06\Mm^{-1}$, so the spherical harmonic
degree would be $\ell\approx k_xR\approx40$,
where $R=700\Mm$ is the solar radius.
On the other hand, as discussed in SBCR, the pressure scale height $H_{\rm d}$
is the only intrinsic length scale in the underlying nonmagnetic
problem, and so the range of dimensionless wavenumbers resolved in our
simulations is
$k_x \gamma H_{\rm d}\approx2$--$15$.
With $\gamma H_{\rm d}\approx0.5\Mm$, this corresponds to values of
$\ell$ that are much larger than those of the pattern seen in \Fig{SM}.
The question is thus, whether in the simulations
a magnetic field with a much larger horizontal
wavelength would still be able to produce signatures that could be discerned
from the diagnostic $k\omega$ diagram.
This is not obvious, given that values of $k_x=0.06\Mm^{-1}$ correspond to
$k_x\gamma H_{\rm d}=0.03$, where with our box geometry we are
unable to produce clear features.
Thus, while it is impossible to make a clear case for helioseismic
applications, our work has opened
the possibility for more targeted searches both theoretically and
observationally.

\section*{Acknowledgements}

Financial support from the Swedish Research Council under the grants
621-2011-5076 and 2012-5797, the European Research Council under the
AstroDyn Research Project 227952 as well as the Research Council of
Norway under the FRINATEK grant 231444 are gratefully acknowledged.
The computations have been carried out at the National Supercomputer
Centres in Link\"oping and Ume{\aa} as well as the Center for Parallel
Computers at the Royal Institute of Technology in Sweden and the Nordic
High Performance Computing Center in Iceland.


\end{document}